\documentclass[%
 reprint,
 amsmath,amssymb,
 aps,
prb,
]{revtex4-2}

\usepackage{graphicx}
\usepackage{dcolumn}
\usepackage{bm}
\usepackage{color}
\usepackage{cancel}
\usepackage{comment}
\usepackage{appendix}
\usepackage{braket}
\allowdisplaybreaks
\usepackage[colorlinks=true,
  allcolors={blue}]{hyperref}
\usepackage{cleveref}

\newcommand{\x}{{\color{black}\langle S^y \rangle}}
\newcommand{\y}{{\color{black}\langle (S^y)^2 \rangle}}
\newcommand{\z}{{\color{black}\langle (S^y)^3 \rangle}}

\begin{document}

\title{Dielectric tunability of magnetic properties \\in orthorhombic ferromagnetic monolayer CrSBr}

\author{Alexander N. Rudenko}
\email{a.rudenko@science.ru.nl}

\author{Malte R\"{o}sner}

\author{Mikhail I. Katsnelson}
\affiliation{\mbox{Radboud University, Institute for Molecules and Materials, Heijendaalseweg 135, 6525AJ Nijmegen, The Netherlands}}

\date{\today}

\begin{abstract}
Monolayer CrSBr is a recently discovered semiconducting spin-3/2 ferromagnet with a Curie temperature around 146\,K. Unlike many other known two-dimensional (2D) magnets, CrSBr has an orthorhombic lattice, giving rise, for instance, to spatial anisotropy of the magnetic excitations within the 2D plane. Theoretical description of CrSBr within a spin Hamiltonian approach turns out to be nontrivial due to the triaxial magnetic anisotropy as well as due to magnetic dipolar interactions, comparable to spin-orbit effects in CrSBr. 
Here, we employ a Green’s function formalism combined with first-principles calculations to systematically study the magnetic properties of monolayer CrSBr in different regimes of surrounding dielectric screening. 
We find that the magnetic anisotropy and thermodynamical properties of CrSBr depend significantly on the Coulomb interaction and its external screening. In the free-standing limit, the system turns out to be close to an easy-plane magnet, whose long-range ordering is partially suppressed. On the contrary, in the regime of large external screening, monolayer CrSBr behaves like an easy-axis ferromagnet with more stable magnetic ordering. 
Despite being relatively large, the magnetic dipolar interactions have only little effect on the magnetic properties.
Our findings suggests that 2D CrSBr is suitable platform for studying the effects of substrate screening on magnetic ordering in low dimensions.
\end{abstract}

\maketitle


\section{Introduction}
Two-dimensional (2D) magnets represent an unique class of materials, which offer great potential for designing spintronic devices with a number of emerging functionalities \cite{Burch2018,Gibertini2019}. Pioneering studies of intrinsic 2D magnets such as CrI$_3$ or Cr$_2$Ge$_2$Te$_6$ have demonstrated rich physics of these materials, opening up new ways for a controllable modification of their properties, which are prospective for various applications \cite{Gong2017,Huang2017,Klein2018}. Experimentally, the tunability of 2D magnets is typically achieved by electrostatic gating \cite{Jiang2018,Huang2018,Jiang2018-2}. Another approaches might include, for example, substrate-induced dielectric screening \cite{Soriano2021}, controllable surface functionalization \cite{Caglayan2022}, and strain engineering \cite{Wu2019,Memarzadeh2021}.

Most of the known van der Waals magnets have a hexagonal crystal structure within the 2D plane, resulting in the isotropic character of their properties at the macroscopic scale. Recently, new types of low-symmetry 2D magnets have been discovered, with CrSBr being a typical representative of this family \cite{Telford2020,Wilson2021}. CrSBr is an orthorhombic van der Waals semiconductor with two inequivalent crystallographic directions for each layer. This gives rise to a strong anisotropy of the electronic and optical properties \cite{Wilson2021,Klein2022,Wu2022}, rendering CrSBr a candidate for studying quasi-1D physics. Further intriguing properties of CrSBr include unusual magneto-electronic coupling \cite{Wilson2021,Telford2022}, nanoscale spin texture engineering \cite{Klein2021}, as well as possible many-body effects \cite{Klein2022}.

Monolayer (ML) CrSBr is a spin-3/2 ferromagnet with the easy axis along the [010] in-plane direction, and an experimentally determined Curie temperature of $\sim$146\,K~\cite{Lee2021}. The ferromagnetism of CrSBr is well understood at the level of the Heisenberg model, as shown by first-principles density functional theory (DFT) calculations, which predict a ferromagnetic exchange coupling between the localized spins at Cr atoms \cite{Hua2020,Yang2021,Hou2022,Esteras2022,Bo2023}, in agreement with the experimental spin-wave spectra \cite{Scheie2022}. With magnetic anisotropy the situation is considerably more involving, as indicated by the conflicting literature reporting different direction of the easy axis in ML-CrSBr \cite{Hua2020,Yang2021,Hou2022}. On the one hand, this inconsistency could be attributed to a relatively small magnetocrystalline anisotropy energy, which is comparable to the magnetic dipole-dipole interactions \cite{Yang2021}, usually ignored in first-principles calculations. On the other hand, the magnetic properties of ML-CrSBr turn out to be highly sensitive to the computational and structural details, such as strain and the Coulomb interaction strength \cite{Esteras2022}. At the same time, the theoretical treatment of low-symmetry magnets is considerably more challenging even at the level of spin Hamiltonians due to the presence of multiple anisotropy terms, becoming especially more complicated if the long-range dipole-dipole interactions are relevant.

Here we systematically study the magnetic properties of ML-CrSBr focusing on the effect of environmental dielectric screening and magnetic dipole-dipole interactions. For this purpose, we use first-principles calculations combined with localized spin models including triaxial magnetic anisotropy, solved by means of Green's function techniques. Despite its highly anisotropic crystal structure, the magnon propagation is weakly anisotropic in ML-CrSBr being almost independent of the dielectric screening. On the contrary, the magnetocrystalline anisotropy is found to be strongly dependent on the Coulomb interaction. Without external screening, the effects of spin-orbit coupling in magnetic anisotropy are small and comparable with the magnetic dipole-dipole interactions. In the presence of external screening, the magnetocrystalline anisotropy is enhanced, leading to a stabilization of the magnetic ordering in ML-CrSBr. For any realistic Coulomb interactions, we always find the easy axis to be along the [010] direction. The Curie temperature is estimated to be around 140-160 K, in good agreement with the experimental data.


The rest of the paper is organized as follows. In Sec.~\ref{sec2}, we provide computational details and the crystal structure of ML-CrSBr. First-principles results on the magnetic anisotropy and the role of magnetic dipolar interaction are discussed in Sec.~\ref{sec3a}. In Sec.~\ref{sec3b}, we estimate the strength of the Coulomb interaction in ML-CrSBr and determine the limits of its tunability by means of external screening. In Sec.~\ref{sec3c}, we present a generalized spin Hamiltonian for ML-CrSBr. The parameters of the spin Hamiltonian determined from first principles are discussed in Secs.~\ref{sec3d} and \ref{sec3e}. In Sec.~\ref{sec3f}, we analyze spin-wave excitations and their dependence on the Coulomb interactions. In Sec.~\ref{sec3g}, we present our results on the temperature-dependent magnetization. In Sec.~\ref{conclusion}, we summarize our results and conclude the paper.

\section{Calculation details\label{sec2}}
\subsection{First-principles calculations}
First-principles calculations were performed using DFT within the projected augmented wave (PAW) formalism \cite{paw1,paw2} as implemented
in the \emph{Vienna ab-initio simulation package} ({\sc vasp}) \cite{Kresse1,Kresse2}.
The exchange-correlation effects were considered within the generalized-gradient approximation (GGA) functional in the Perdew-Burke-Ernzerhof parametrization \cite{pbe}. To account for the on-site Coulomb repulsion within the 3$d$ shell of Cr atoms, we used a simplified version of the DFT+$U$ scheme \cite{Dudarev1998} with the effective Coulomb interaction $U_{\mathrm{eff}}=U-J_H$, where $J_H$ is the Hund's exchange interaction.
A 400 eV energy cutoff for the plane-waves and a convergence threshold of $10^{-8}$ eV were used. All calculations were performed using a ($2\times 2$) supercell containing 6 Cr atoms.
A vacuum layer of 15 \AA~was introduced in the direction perpendicular to the ML-CrSBr surface to eliminate spurious interactions between the supercell images.
The Brillouin zone was sampled by a (6$\times$4) ${\bf k}$-point mesh. 
The spin-orbit coupling (SOC) was treated perturbatively \cite{vasp_soc}.

The Coulomb interaction strength is estimated within the constrained Random Phase Approximation (cRPA) scheme \cite{aryasetiawan_frequency-dependent_2004} allowing us to calculate all static matrix elements $U_{ijkl} = \braket{ w_i w_j | \mathcal{U} | w_k w_l }$ within a Wannier orbital basis describing the Cr $d$ states. For the Wannierization we utilize a spin-unpolarized GGA DFT band structure in which the half-filled Cr $d$ states are clearly disentangled from all other bands and which we project to Cr-centered $d$ orbitals with rotated local bases. To suppress any metallic screening from the Cr $d$ states, we exclude them from screening within the cRPA calculations. For these calculations we use the primitive unit cell in the $U_\text{eff}=0$ crystal structure, (16$\times$16) ${\bf k}$-point meshes, and a vacuum separation between periodically repeated slabs in $z$ direction of $25\,$\AA. All cRPA calculations are performed within {\sc vasp} using algorithms implemented by M. Kaltak~\cite{kaltak_merging_2015}.

\subsection{Crystal structure}

\begin{figure}[tbp]
\centering
\includegraphics[width=0.75\linewidth]{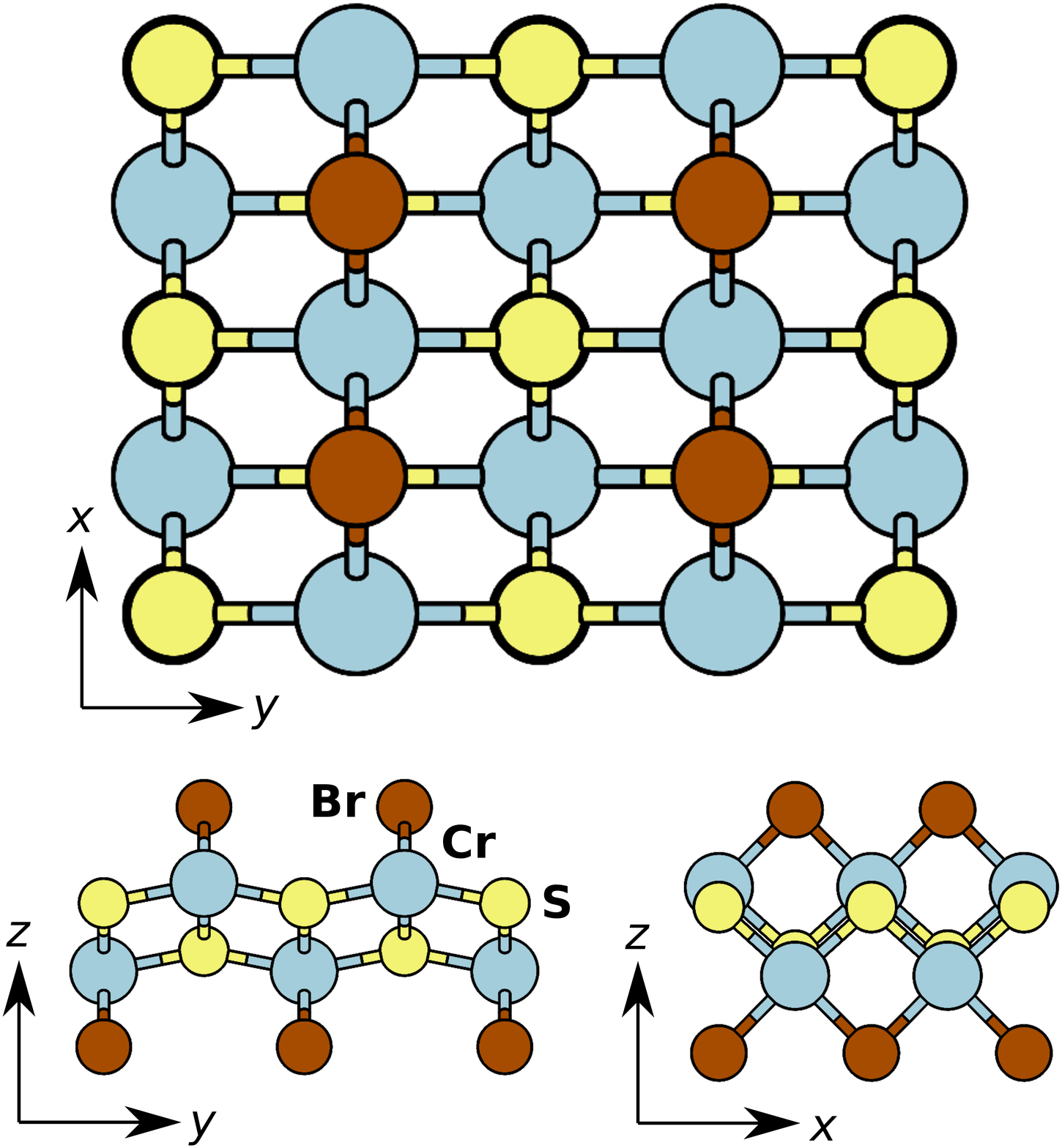}
\caption{Schematic crystal structure of monolayer CrSBr shown in three different projections. Brown, blue, and yellow balls correspond to Br, Cr, and S atoms, respectively.}
\label{structure}
\end{figure}

ML-CrSBr has an orthorhombic crystal structure with two distinct in-plane crystallographic directions, as shown schematically in Fig.~\ref{structure}. The monolayer structure is centrosymmetric with a point group symmetry $D_{2h}$. The Cr atoms reside in a distorted octahedral coordination formed by S and Br atoms. Along the [100] direction ($x$),  the Cr atoms are connected to the neighboring Cr atoms by S and Br atoms, forming $\sim$90$^\mathrm{o}$ bonds. Along the [010] direction ($y$), Cr atoms are connected only by the S atoms with the bond angle around 180$^\mathrm{o}$. In our calculations, we use two set of lattice constants: (i) Optimized lattice constants obtained without Coulomb corrections ($U_\text{eff}=0$), $a=3.54$ \AA~and $b=4.74$ \AA, which are close to the experimental constants of bulk CrSBr $a_{\mathrm{exp}}=3.51$ \AA~and $b_{\mathrm{exp}}=4.77$~\AA~\cite{GOSER1990129}; (ii) Optimized lattice constants obtained for each specific $U_\text{eff}$ considered, which are up to 2-3\% larger compared to the optimization with $U_\text{eff}=0$. In what follows, our results are presented for the two cases separately.

\section{Results}

\subsection{Magnetic anisotropy and the role of magnetic dipolar interactions\label{sec3a}}

\begin{figure}[tbp]
\centering
\includegraphics[width=1.00\linewidth]{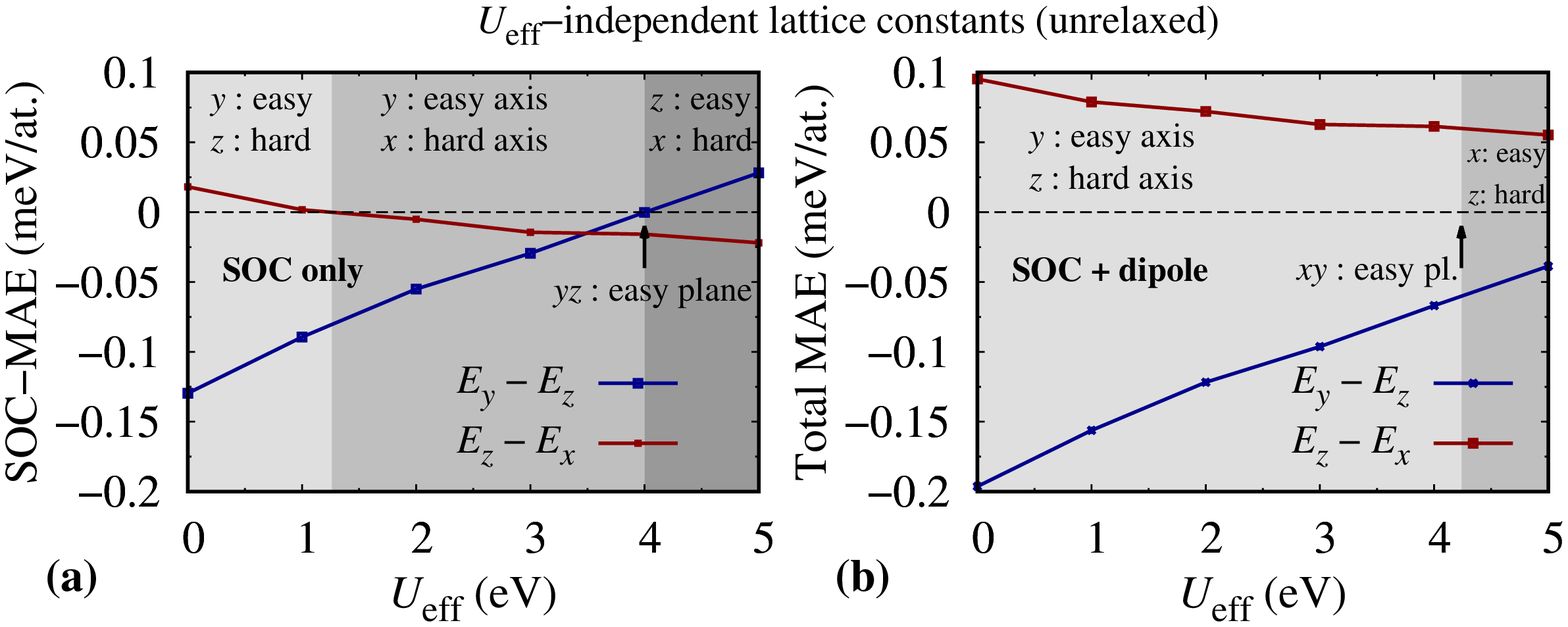}
\includegraphics[width=1.00\linewidth]{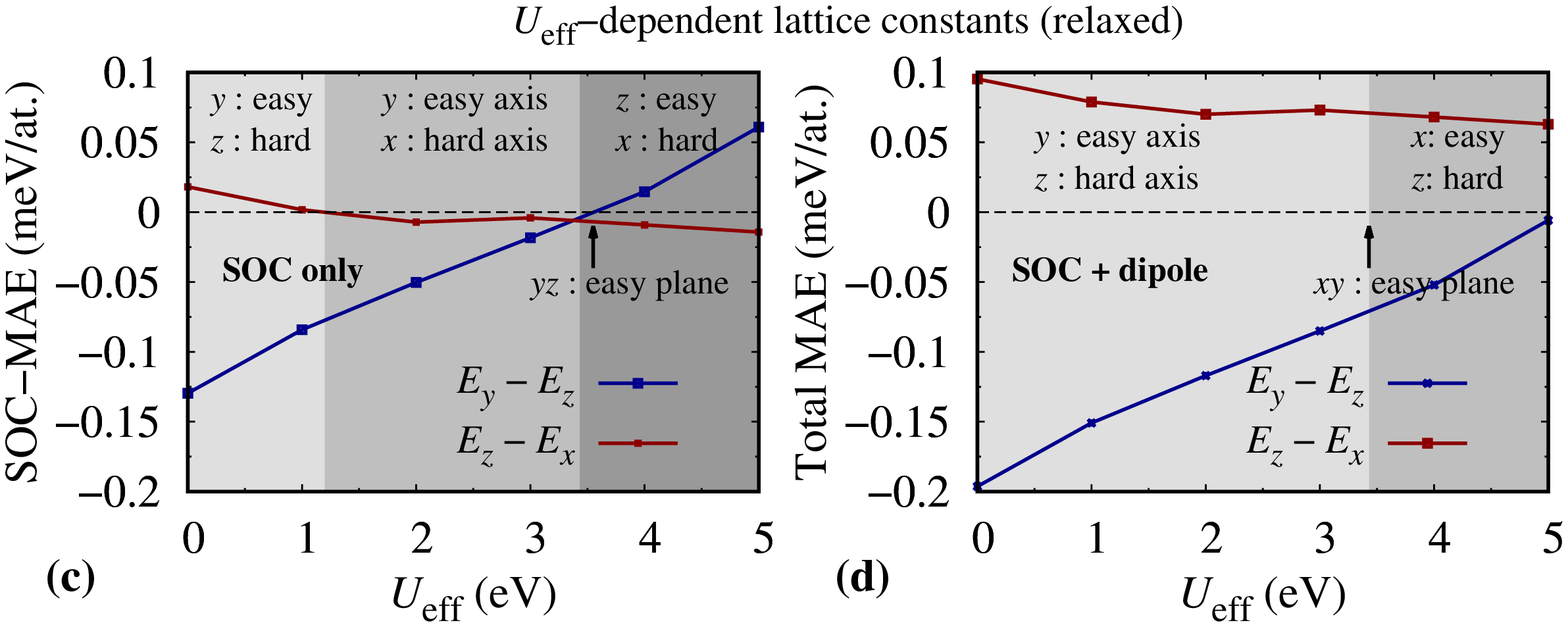}
\includegraphics[width=1.00\linewidth]{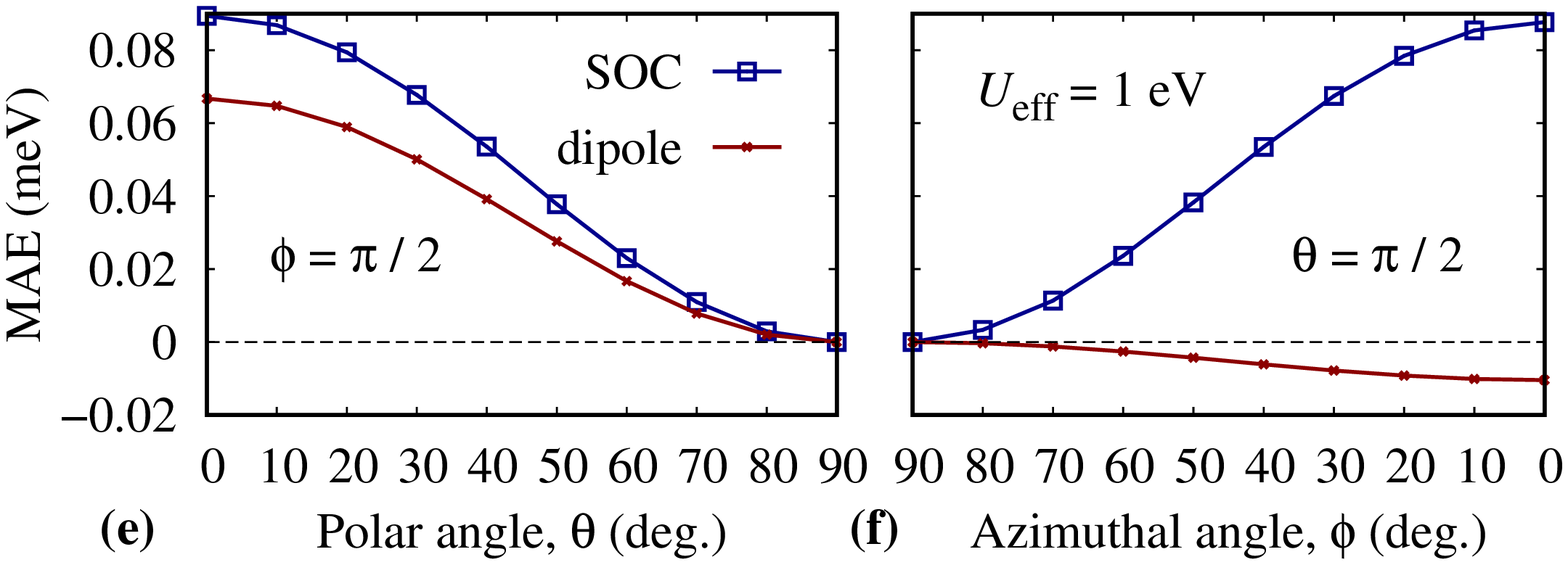}
\caption{Magnetic anisotropy energies calculated in ML-CrSBr as a function of the Coulomb interaction $U_\text{eff}$. 
Left panels (a) and (c) show SOC contribution to MAE only without dipolar interactions considered, while in the right panels (b) and (d) the total MAE is presented. Upper panels (a) correspond to the $U_\text{eff}$-independent lattice constants used in the calculations, while the lower panels (c) and (d) include $U_\text{eff}$-induced relaxation effects into account. The arrows indicate the $U_\text{eff}$ values at which the system turns into an easy-plane ferromagnet. Shaded areas correspond to the regions with different easy/hard axis. (e) and (f) show dependence of the SOC and dipolar contributions to MAE on the polar $\theta$ and azimuthal $\phi$ angles for the unrelaxed case with $U=1$ eV.}
\label{anisotropy}
\end{figure}

We first analyze the magnetic anisotropy energy in ML-CrSBr, which demonstrates a remarkable behavior compared to other 2D magnets. Figure \ref{anisotropy}(a) shows the SOC contribution to the two components of MAE, namely, $E_y-E_z$ and $E_z-E_x$ as a function of $U_\text{eff}$. The results allow us to distinguish between the three different ground state magnetic configurations. Up to $U_\text{eff}\approx4$ eV, the $y$ direction  corresponds to the easy axis, while the hard axis changes from $z$ to $x$ at $U_\text{eff}\gtrsim 1$ eV. As $U_\text{eff}$ increases, the magnetization along $y$ becomes less  favorable, reaching the crossover point at $U_\text{eff}\approx4$ eV, after which the easy-axis becomes oriented along $z$. 

The situation becomes considerably different upon taking the dipole-dipole interaction into account, see Fig.~\ref{anisotropy}(b). In this case, the out-of-plane direction $z$ is highly unfavorable, so that $z$ always corresponds to the hard axis, independently of $U_\text{eff}$. At the same time, the dipolar interaction tends to align spins along the $x$ axis because the corresponding lattice constant $a$ is the smallest. As a result, at sufficiently large $U_\text{eff}$, where the SOC contribution to MAE is low, the $x$ direction becomes the easy axis of ML-CrSBr. 

The relaxation of the lattice parameters in the presence of $U_\text{eff}$ does not lead to any qualitative effects, as one can see from Figs.~\ref{anisotropy}(c) and (d). Quantitatively, the crossover points shift toward lower $U_\text{eff}$ values if the relaxation is taken into account. Particularly, the easy plane transition point (marked by arrows in Fig.~\ref{anisotropy}) is now at $U_\text{eff}\approx 3.5$ eV, both for the SOC-only contribution and for the total MAE. 

In Figs.~\ref{anisotropy}(e) and (f), we show the angular dependence of the SOC and dipolar contributions to the magnetic energy calculated for $U_\text{eff}=1$ eV. Similar to the dipolar contribution, one can clearly see that the SOC contribution follows a $\sim\mathrm{cos}^2\theta$ ($\mathrm{cos}^2\phi$) behavior, suggesting that the quadratic anisotropy terms are sufficient for the construction of the spin Hamiltonian for ML-CrSBr.

\subsection{Coulomb interaction and dielectric sceeening\label{sec3b}}
Up to now, we did not specify the strength of the Coulomb interaction and considered $U_\text{eff}$ as a parameter. In a real situation, $U_\text{eff}$ is determined by the environmental conditions such as the external dielectric screening governed, for example, but the underlying substrates. In this section, we estimate $U_\text{eff}$ of free-standing ML-CrSBr and determine the degree of its tunability by means of dielectric environments. To this end, we perform constrained random phase approximation (cRPA) calculations and investigate the Cr $d$ local intra- and inter-orbital density-density matrix elements $U = \frac{1}{5}\sum_{i}U_{iiii}$ and $U' = \frac{1}{20}\sum_{i \neq j}U_{ijji}$, respectively, as well as the averaged Hund's exchange elements $J_H = \frac{1}{20}\sum_{i \neq j}U_{ijij}$. To investigate the influence of the environmental screening we use our Wannier Function Continuum Electrostatics (WFCE) approach~\cite{rosner_wannier_2015} to calculate $U$, $U'$ and $J_H$ as a function of $\varepsilon_{env}$ referring to the screening from dielectric encapsulation. For the free-standing layer, i.e. $\varepsilon_{env} = 1$ we find $U\approx 3.68\,$eV, $U' \approx 2.86\,$eV, and $J_H \approx 0.39 \,\text{eV} \approx (U-U')/2$ showing the approximate rotational-invariance of the Coulomb tensor. A realistic value for our DFT$+U$ calculation for the free-standing monolayer is thus $U_\text{eff}=U-J_H=3.28\,$eV. The dielectric environmental screening strongly reduces density-density interactions, while $J_H$ is barely affected. This is a result of the mono-pole character of the environmental screening model we apply here and which we previously benchmarked for CrI$_3$ by means of full cRPA calculations explicitly taking environmental screening into account~\cite{Soriano2021}. In Table~\ref{tab:U}, we show the resulting averaged matrix elements and find that $U$ and $U'$ can be both reduced by about $1\,$eV by dielectric environments with $\varepsilon_{env} < 10$, while high-$k$ dielectrics might reduce $U$ and $U'$ even up to $1.8\,$eV. As the Hund's exchange is not affected, $U_\text{eff}$ is thus tunable on the same range.

\begin{table}[tbp]
    \caption{WFCE+cRPA averaged local Cr $d$ Coulomb matrix elements as a function of the screening from a dielectric encapsulation of the CrSBr monolayer. The values for $\varepsilon_{env}=\infty$ are obtained using the Richardson extrapolation method.}
    \label{tab:U}
    \centering
    \begin{tabular}{c|c|c|c|c}
        $\varepsilon_{env}$ & $U$ (eV) & $U'$ (eV) & $J_H$ (eV) & $U_\text{eff} = U-J_H$ (eV) \\
        \hline\hline
            1 &    3.68 &  2.86 &  0.39 &  3.28 \\
            2 &    3.15 &  2.34 &  0.39 &  2.76 \\
            4 &    2.74 &  1.93 &  0.39 &  2.35 \\
            8 &    2.43 &  1.61 &  0.39 &  2.03 \\
           16 &    2.20 &  1.39 &  0.39 &  1.81 \\
           32 &    2.06 &  1.25 &  0.39 &  1.67 \\
           64 &    1.98 &  1.17 &  0.39 &  1.58 \\
       $\infty$\footnote{Extrapolated values.} &    1.90 &  1.08 &  0.39 &  1.40 \\
           \hline
    \end{tabular}
\end{table}


\subsection{Spin Hamiltonian\label{sec3c}}
For orthorhombic magnetic crystals with inversion symmetry, the most general form of the quadratic spin Hamiltonian can be written as
\begin{equation}
\label{hamilt}
H = H_{0} + H_{SIA} + H_{AE} + H_{D},
\end{equation}
where
\begin{equation}
\label{H0}    
H_0=\sum_{ij}J_{ij}{\bf S}_i{\bf S}_j
\end{equation}
is the Heisenberg term with $J_{ij}$ being the isotropic exchange interaction between lattice sites $i$ and $j$ with spins $S_i$ and $S_j$, 
\begin{equation}
\label{HSIA}
H_{SIA} = D\sum_{i}(S^y_i)^2 + E\sum_{i}\left[(S^z_i)^2-(S^x_i)^2\right]
\end{equation}
describes single-ion anisotropy (SIA) arising from the spin-orbit coupling (SOC) and characterized by the parameters $D$ and $E$,
\begin{equation}
\label{HAE}
H_{AE} = \sum_{ij}K_{ij}S_i^yS_j^y + \sum_{ij}\Gamma_{ij}(S_i^zS_j^z - S_i^xS_j^x)
\end{equation}
is the (symmetric) anisotropic exchange interaction between the sites $i$ and $j$ controlled by the matrix elements $K_{ij}$ and $\Gamma_{ij}$. Finally,
\begin{equation}
\label{HD}
H_{D} = \frac{\Omega}{2}\sum_{ij}\frac{1}{|{\bf R}_{ij}|^3}\left({\bf S}_i{\bf S}_j - 3\frac{({\bf S}_i\cdot {\bf R}_{ij})({\bf S}_j\cdot {\bf R}_{ij})}{{\bf R}_{ij}^2}\right)
\end{equation}
is the dipolar interaction with ${\bf R}_{ij}={\bf R}_i-{\bf R}_j$ being the lattice vector connecting the sites $i$ and $j$, and $\Omega=g^2\mu_0\mu_B^2/4\pi$ is the dipole-dipole interaction constant where $g\approx2$ is the $g$-factor. In what follows, we consider the situation in which $y$ is the spin quantization axis, and $z$ is the direction perpendicular to the 2D plane of a crystal, such that the vectors ${\bf R}_{ij}$ are mostly confined in the $xy$ plane.

To determine the parameters entering Eqs.~(\ref{H0}), 
 (\ref{HSIA}), and (\ref{HAE}) for ML-CrSBr, we construct a series of collinear magnetic configurations and calculate their energies using DFT taking SOC into account. To this end, we consider a ($2\times2$) supercell and determine the exchange parameters up to the fourth nearest neighbor. In total, we consider 15 inequivalent magnetic configurations, allowing us to estimate 14 parameters which determine the spin Hamiltonian (see Appendix \ref{appendix1} for the explicit expressions). 

\subsection{Isotropic exchange interactions\label{sec3d}}

\begin{figure}[tb]
\centering
\includegraphics[width=1.05\linewidth]{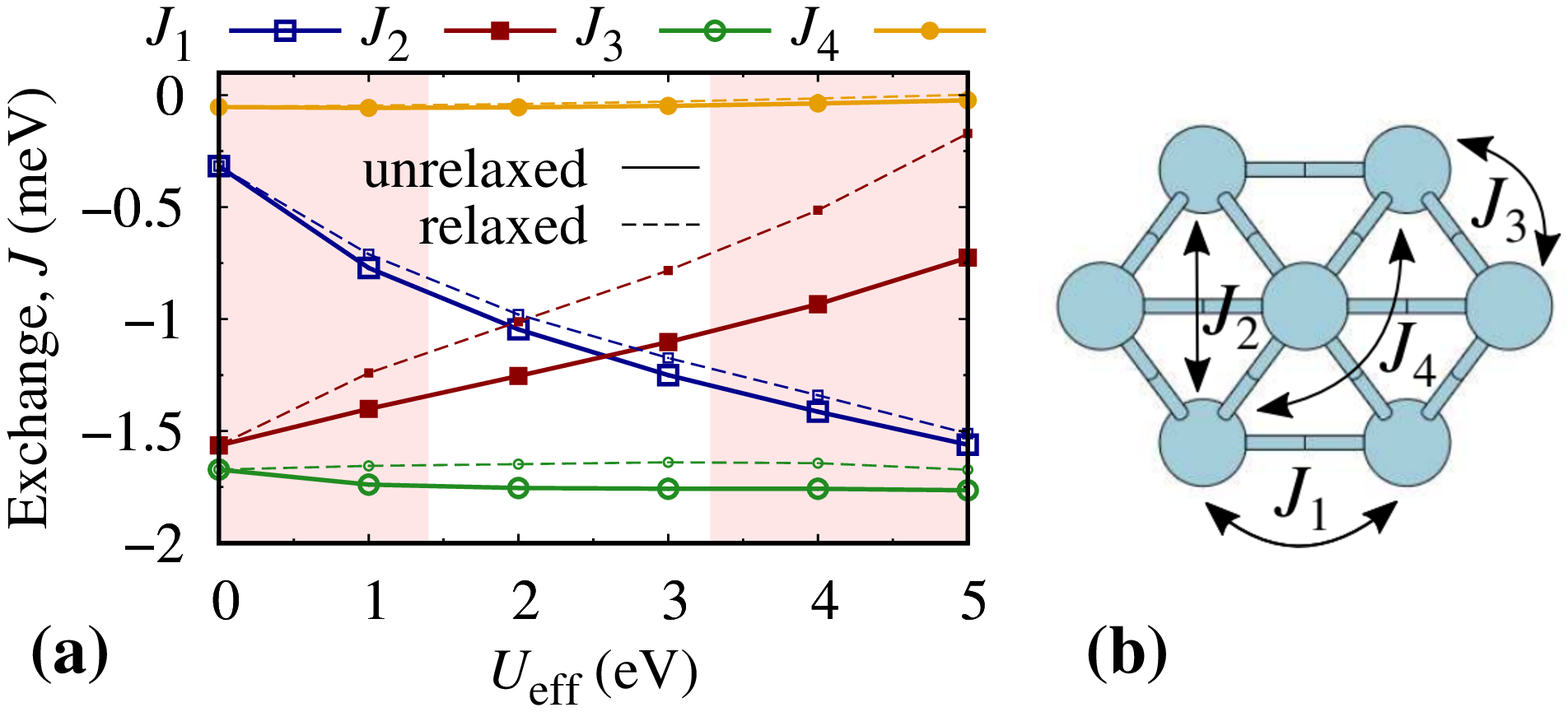}
\includegraphics[width=0.67\linewidth]{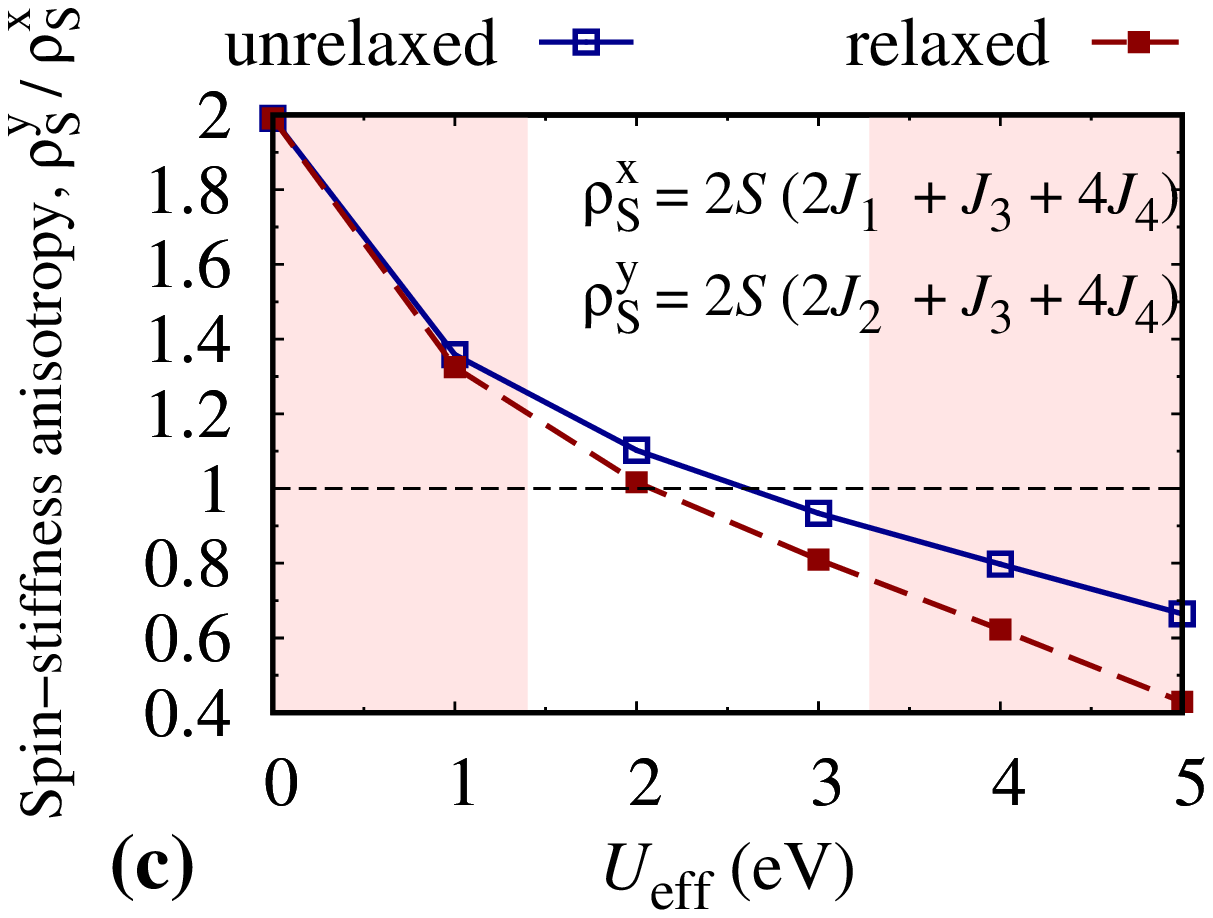}
\caption{(a) Isotropic exchange interactions $J$ shown as a function of the Coulomb interaction $U_\text{eff}$ calculated in ML-CrSBr for the $U_\text{eff}$-independent (unrelaxed) and $U_\text{eff}$-dependent (relaxed) geometries. (b) Schematic representation of the spin lattice and the relevant exchange interactions. (c) The anisotropy ratio of the spin-stiffness constant $\rho^y_S/\rho^x_S$ shown as a function of $U_\text{eff}$. The unshaded (white) region corresponds to the realistic values of $U_\text{eff}$ estimated in Sec.~\ref{sec3b}.}
\label{exchange}
\end{figure}

 Figure~\ref{exchange}(a) shows the calculated isotropic exchange interaction for the four nearest neighbors as a function of the Coulomb interaction $U_\text{eff}$. In order to capture the effect of the $U_\text{eff}$-dependent lattice constants, we also show the results obtained when the structure was fully relaxed for each $U_\text{eff}$ considered [dashed lines in Fig.~\ref{exchange}(a)]. From Fig.~\ref{exchange}(a) one can see that all the exchange interactions are ferromagnetic, with the dominant contribution coming from the three nearest neighbor interactions $J_1$, $J_2$, and $J_3$ [see Fig.~\ref{exchange}(b) for notation], which are of the order of 1 meV. More distant couplings (e.g., $J_4$) are substantially smaller, and can thus be neglected in practical calculations. The nearest neighbor exchange $J_3$ is virtually independent of $U_\text{eff}$, whereas $J_1$ and $J_2$ exhibit a pronounced dependence. While the interaction between the spins along the $x$ direction ($J_1$) increases with $U_\text{eff}$, the interaction along the $y$ direction ($J_2$) shows an opposite tendency. This behavior suggest that the spin excitations in ML-CrSBr are spatially anisotropic. Interestingly, there is a crossing point between $J_1$ and $J_2$, at which the isotopic behavior is restored. The effect of structural relaxation in the presence of additional Coulomb repulsion between the Cr $d$ electrons is a slight lattice expansion, which at $U_\text{eff}=3$ eV is around 1\% and 2\% for the $a$ and $b$ lattice constants, respectively. This lattice expansion leads to a reduction of the exchange interaction, which is clearly seen in Fig.~\ref{exchange}(a). Moreover, as the $b$ lattice constant is more sensitive to $U_\text{eff}$, the difference between the relaxed and unrelaxed exchange in the corresponding direction ($J_2$) is more pronounced.

 Let us now analyze the effect of the spatial anisotropy on the spin-wave dispersion. For this purpose, we first ignore the magnetic anisotropy in the spin Hamiltonian Eq.~(\ref{hamilt}), and focus on the low-energy excitations. At $T=0$ the isotropic Hamiltonian can be transformed to a diagonal form (e.g., using a Holstein-Primakoff transformation), yielding the following spin-wave Hamiltonian for two equivalent sublattices \cite{Rusz2005}:
\begin{equation}
\label{H0_q}
H^{mn}_0({\bf q}) = 2S\left[\sum_{p}J_{mp}({\bf 0})\right]\delta_{mn} - 2SJ_{mn}({\bf q}),
\end{equation}
where $m$, $n$, and $p$ are sublattice indices, and ${\bf q}$ is the wave vector. Here, $J_{mn}({\bf q})=\sum_{\bf R}e^{-i{\bf q}\cdot{\bf R}}J_{mn}({\bf R})$ with ${\bf R}$ being a vector connecting the lattice sites. Also, $J_{11}({\bf q})=J_{22}({\bf q})$ and $J_{12}({\bf q})=J^*_{21}(\bf q)$.
Keeping only four nearest-neighbors, for an orthorhombic 2D crystal we can explicitly write
\begin{multline}
    J_{11}({\bf q}) = 2J_1\,\mathrm{cos}(q_xa) + 2J_2\,\mathrm{cos}(q_yb) + \\ 
    4J_4\,\left[\mathrm{cos}(q_xa) \, \mathrm{cos}(q_yb)\right],
\end{multline}
and
\begin{equation}
    J_{12}({\bf q}) = 4J_3\,\mathrm{cos}\left(\frac{q_xa}{2}\right) \mathrm{cos}\left(\frac{q_yb}{2}\right),
\end{equation}
where $a$ and $b$ are the lattice parameters, and $q_x$ ($q_y$) are the wave vectors ranging from 0 to 2$\pi/a$ (2$\pi/b$). 
Diagonalizing Eq.~(\ref{H0_q}), we obtain the following spin-wave dispersion

\begin{equation}
\label{w0}
\omega^{\pm}_0({\bf q}) = 2S\left[ J_{11}({\bf 0}) + J_{12}({\bf 0}) - J_{11}({\bf q}) \pm J_{12}({\bf q}) \right].
\end{equation}
Expanding the lower branch at ${\bf q}\rightarrow 0$, we 
arrive at $\omega^{-}_0({\bf q}) = \rho_S^x q_x^2 + \rho_S^y q_y^2$, where $\rho_S^x=2S(2J_1+J_3+4J_4)$ and $\rho_S^y=2S(2J_2+J_3+4J_4)$ are the spin-stiffness constants in the $x$ and $y$ directions, respectively. One can see that the spin-stiffness is anisotropic provided that $J_1$ and $J_2$ are different. In Fig.~\ref{exchange}(c), we show the spin-stiffness anisotropy $\rho_S^y/\rho_S^x$ calculated as a function of $U_\text{eff}$ using the exchange interactions from Fig.~\ref{exchange}(a). As expected from the strong spatial anisotropy of the exchange constants $J_1$ and $J_2$ in the limit of zero Coulomb interactions, the spin-stiffness anisotropy of around $2.0$ is observed at $U_\text{eff}=0$. At larger $U_\text{eff}$, the anisotropy becomes smaller, with a crossover point around $U_\text{eff}=2$ eV. Therefore, our results demonstrate that ML-CrSBr has a preferred direction of the magnon propagation, which is expected to be dependent on the environmental conditions such as external dielectric screening.

\subsection{Anisotropic exchange and single-ion anisotropy\label{sec3e}}
\begin{figure}[!tbp]
\centering
\includegraphics[width=1.05\linewidth]{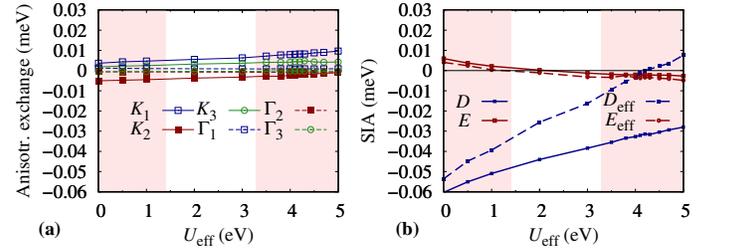}
\caption{(a) Anisotropic exchange parameters [see Fig.~\ref{exchange}(b) for notation] and (b) single-ion anisotropy entering the spin Hamiltonian Eq.~(\ref{hamilt}) for ML-CrSBr shown as a function of $U_\text{eff}$. The scale of the two plots is made intentionally the same, in order to demonstrate relative strength of the two effects. The effective parameters $D_{\mathrm{eff}}$ and $E_{\mathrm{eff}}$ are obtained by neglecting the anisotropic exchange terms Eq.~(\ref{HAE}) in the spin Hamiltonian. The unshaded (white) region corresponds to the realistic values of $U_\text{eff}$ estimated in Sec.~\ref{sec3b}. The results are shown for the $U_\text{eff}$-independent (unrelaxed) lattice constants.}
\label{GKDE}
\end{figure}

The calculated single-ion anisotropy (\ref{HSIA}) 
 and anisotropic exchange parameters (\ref{HAE}) are shown in Fig.~\ref{GKDE} for the case of $U_\text{eff}$-independent lattice constants. The anisotropic exchange in ML-CrSBr is extremely small, being of the order of $\mu eV$. On the other hand, the single-ion anisotropy parameters $D$ and $E$ are larger by 1--2 orders of magnitude, suggesting that the effects of the anisotropic exchange can be safely neglected. In what follows, we exclude the anisotropic exchange term [Eq.~(\ref{HAE})] from the consideration, and recalculate the effective SIA parameters $D_{\mathrm{eff}}$ and $E_{\mathrm{eff}}$, allowing us to quantitatively describe MAE presented in Fig.~\ref{anisotropy}. As a result of this simplification, no essential changes neither in the spin-wave excitations nor in the thermodynamic behavior of ML-CrSBr are expected.

 Having determined the parameters of the spin Hamiltonian, it is worth noting that in the regime of small $U_\text{eff}$, the system can be treated as an easy-axis ferromagnet with SIA and negligible dipole-dipole interactions. In this situation, the corresponding spin Hamiltonian can be solved by means conventional methods such as Green's function techniques \cite{Valkov1982,Frobrich2000,Frobrich2006} or self-consistent spin-wave theories \cite{Irkhin1999}. In the presence of dipole-dipole interactions, extensions of these methods are available \cite{Bruno1991,Frobrich2001,Grechnev2005}. At lager $U_\text{eff}$, when the parameters $D_{\mathrm{eff}}$ and $E_{\mathrm{eff}}$ are comparable in magnitude,
the system is close to an easy-plane ferromagnet. This situation is considerably more complicated due to the effects related to mixing of the eigenstates of the $S^y$ operator \cite{Hu1999}.

\subsection{Spin-wave excitations\label{sec3f}}
Let us now consider spin-wave excitations of ML-CrSBr. For this purpose, we closely follow the Green's function approach formulated in Ref.~\cite{Hu1999} for easy-plane ferromagnets, whose generalization for magnets with triaxial symmetries and multiple equivalent sublattices is straightforward.

For the system under consideration, the magnon dispersion relation can be written as \begin{equation}
\label{dispersion}
    E^{\pm}({\bf q}) = \sqrt{[F^{\pm}_1({\bf q})]^2 - [F_2({\bf q})]^2},
\end{equation}
where
\begin{multline}
\label{f1}
    F^{\pm}_1({\bf q}) = \frac{\langle S^y\rangle}{S} \omega_0^{\pm}({\bf q}) + \frac{\Omega}{2}\langle S^y \rangle p_{zz}({\bf q}) + \frac{\Omega}{2}\langle S^y \rangle p_{xx}({\bf q}) \\
    - \Omega \langle S^y \rangle p_{yy}({\bf 0}) + 2D\Phi\langle S^y \rangle,
\end{multline}
\begin{equation}
\label{f2}
 F_2({\bf q}) = \frac{\Omega}{2}\langle S^y \rangle p_{zz}({\bf q}) - \frac{\Omega}{2}\langle S^y \rangle p_{xx}({\bf q})  + 2E\Phi\langle S^y \rangle.
\end{equation}
Here, $\omega_0^{\pm}$({\bf q}) is the zero-temperature isotropic contribution to the dispersion relation [see Eq.~(\ref{w0})], $\langle ... \rangle = \mathrm{Tr}(...e^{\beta H})/\mathrm{Tr}(e^{-\beta H})$ is the ensemble average with $\beta=1/k_BT$. $p_{\alpha \alpha}({\bf q})$ is the Fourier transform of the dipole-dipole interaction energy per spin for the case when the magnetization is aligned along the $\alpha$ direction, i.e.
 \begin{equation}
     p_{\alpha\alpha}({\bf q}) = \sum_{{\bf R}_{0i}} \frac{1}{|{\bf R}_{0i}|^3}\left( 1 - 3\frac{R^\alpha_{0i}R^\alpha_{0i}}{{\bf R}^2_{0i}} \right) e^{i{\bf q}{\bf R}_{0i}}.          
 \end{equation}
In Eqs.~(\ref{f1}) and (\ref{f2}), we assume that $y$ is the spin quantization axis, and $z$ is the direction perpendicular to the surface of the material.

Equation~(\ref{dispersion}) is obtained by means of the Green's function technique with the Tyablikov decoupling \cite{Tyablikov} for the intersite spin operators $S^y_i(t)S^{\pm}_j(t) \rightarrow \langle S^y \rangle S_j^{\pm}$ ($i\neq j$), and the Anderson-Callen decoupling \cite{AndersonCallen} for the on-site spin operators $S^y_i(t)S^{\pm}_i(t) + S^{\pm}_i(t)S^y_i(t) \rightarrow 2\Phi \langle S^y \rangle S_i^{\pm}$,
where
\begin{equation}
    \Phi = 1 - \frac{1}{2S^2}[S(S+1) - \langle (S^y)^2\rangle]
    \label{decoupling}
\end{equation}
is the decoupling function, which satisfies the kinematic condition, i.e. $\Phi=0$ for $S=1/2$. In particular, for an easy-axis ferromagnet ($E=0$) without dipolar interactions ($\Omega=0$), Eq.~(\ref{dispersion}) simplifies to $E^{\pm}({\bf q}) = \frac{\langle S^y\rangle}{S}\omega_0^{\pm}({\bf q}) + 2D\Phi \langle S^y \rangle$, which at $T\rightarrow0$ takes the well-known form $E^{\pm}({\bf q}) = \omega^{\pm}_0({\bf q}) + (2S-1)D$ \cite{Balucani1979}.

\begin{figure}[t]
\centering
\includegraphics[width=1.05\linewidth]{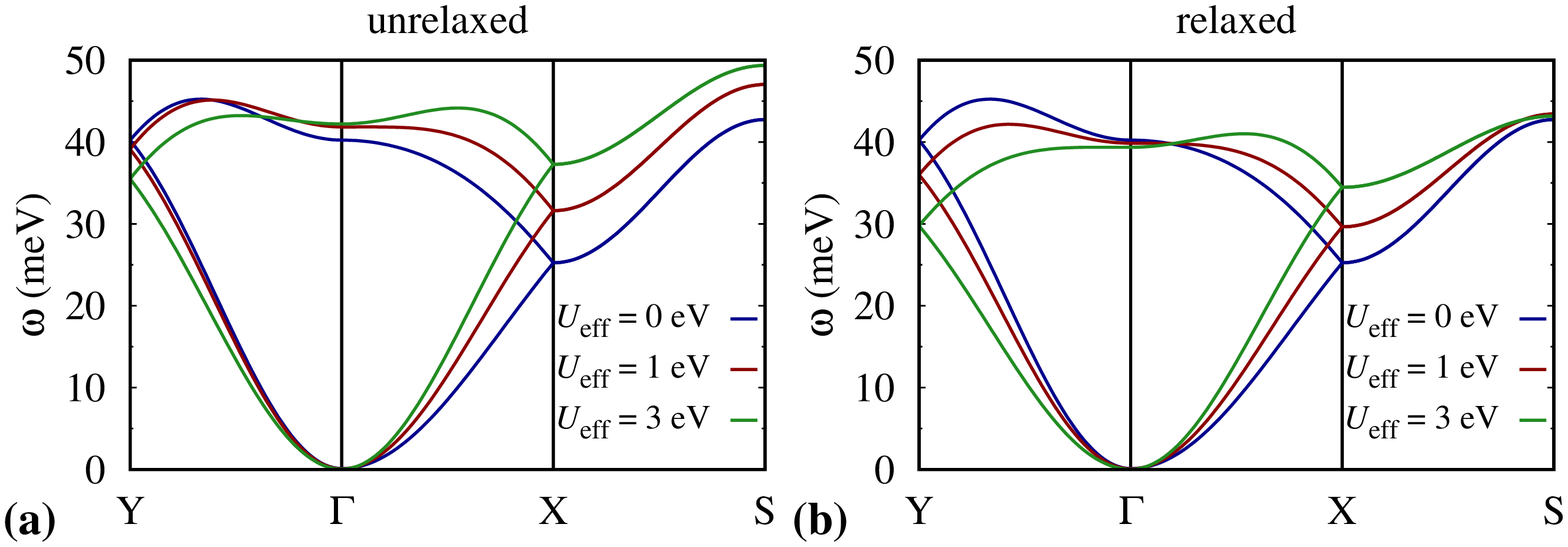}
\includegraphics[width=1.05\linewidth]{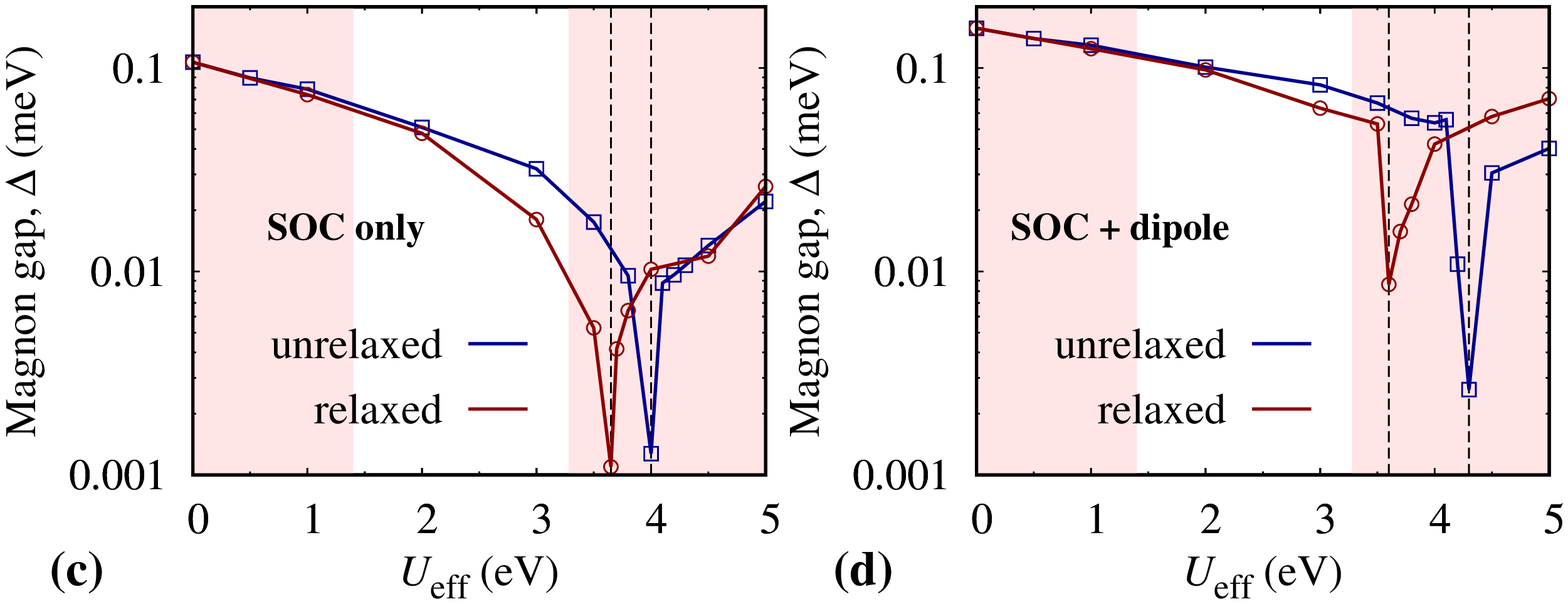}
\caption{Top panels: Magnon dispersion relation calculated for different values of $U_\text{eff}$ in ML-CrSBr using (a) $U_\text{eff}$-independent lattice constants and (b) $U_\text{eff}$-dependent lattice constants. Bottom panels: Magnon energy gap $\Delta$ at the $\Gamma$ point shown as a function of $U_\text{eff}$. (c) shows the SOC-only contribution to the magnon gap, while (d) also takes dipolar interactions into account. Vertical dashed lines correspond to the transition between the easy axes. The unshaded (white) region in (c) and (d) corresponds to the realistic values of $U_\text{eff}$ estimated in Sec.~\ref{sec3b}. }
\label{magnon_dispersion}
\end{figure}

Figures \ref{magnon_dispersion}(a) and \ref{magnon_dispersion}(b) show the magnon dispersion of ML-CrSBr calculated at $T=0$ for different values of $U_\text{eff}$ using $U_\text{eff}$-independent and $U_\text{eff}$-dependent lattice constants, respectively. In all cases, the highest excitation energy is around 40--50 meV, which is in agreement with experimental results from inelastic neutron scattering \cite{Scheie2022}.
In accordance with the previously calculated spin-stiffness, one can see a difference in the magnon dispersion along the $\Gamma$--X and $\Gamma$--Y, which becomes less pronounced as $U_\text{eff}$ increases. At a large energy scale, the dispersion is quadratic around the $\Gamma$ point, which is typical for an easy-axis ferromagnet. The variation of the dispersion relation with $U_\text{eff}$ is primarily related to the variation of the isotropic exchange parameters. At a small scale, some deviation from the quadratic behavior could be observed close to the transition points, where easy axis switching takes place. As the region of the $U_\text{eff}$ parameters, in which a linear dispersion (typical for an easy-plane ferromagnet) is realized, is extremely narrow, it is unlikely that it could be observed experimentally. On the other hand, from Figs.~\ref{magnon_dispersion}(c) and \ref{magnon_dispersion}(d) one can see that the magnon gap varies dramatically (note the logarithmic scale) with $U_\text{eff}$, being a consequence of the single-ion anisotropy which is strongly dependent on $U_\text{eff}$. The inclusion of the dipolar interaction suppresses the reduction of the gap. Although the dipolar interaction does not provide a contribution to the gap for systems that are isotropic within the 2D plane \cite{Hu1999}, the orthorhombic symmetry of ML-CrSBr ensures a non-zero magnon gap even in the absence of SOC-induced anisotropy. A strong variation of the magnon gap with $U_\text{eff}$ in 2D suggests that the Coulomb interaction plays a nontrivial role in the thermodynamics of ML-CrSBr and that substrate screening can thus have a significant impact.

\subsection{Thermodynamical properties\label{sec3g}}
To calculate the thermodynamical properties of ML-CrSBr, we use the Green's functions formalism. For convenience, we rewrite the single-ion contribution to the Hamiltonian (\ref{HSIA}) as
\begin{equation}
    H_{SIA} = D\sum_i(S_i^y)^2 + \frac{E}{2}\sum_i \left[ (S^+_i)^2 + (S^-_i)^2 \right],
    \label{HSIA_ladder}
\end{equation}
where $S^+$ and $S^-$ are the ladder operators.
The presence of the $S^+$ and $S^-$ terms create non-diagonal elements in the Hamiltonian (\ref{HSIA_ladder}) in the basis of the $S_i^y$ eigenvectors, which render the treatment of this situation more complicated compared to the easy axis case, with $E=0$ and $D<0$. In order to solve the general Hamiltonian (\ref{hamilt}), we employ the following two kinds of the Green's functions \cite{Hu1999}
\begin{equation}
    G^{(n)}_{1,2} = -i \Theta (t-t') \langle [A_i^{1,2}(t), B_j(t')] \rangle,
\end{equation}
where $A_i^{1}(t) = S_i^+$, $A_i^{2}(t) = S_i^-$, and $B_j(t') = (S_j^-(t'))^n (S_j^+(t'))^{n-1}$. Here, $n=1,2, ..., 2S$, and $A(t) = e^{iHt} A e^{-iHt}$. With this definition, the equations of motion for the Green's functions can be written using the transformations given prior to Eq.~(\ref{decoupling}) as
\begin{multline}
[\omega - F_1^{\pm}({\bf q})]G_1^{(n)}({\bf q},\omega)
- F_2({\bf q}) G_2^{(n)}({\bf q},\omega) = \langle g_1^{(n)} \rangle \\
[\omega + F_1^{\pm}({\bf q})]G_2^{(n)}({\bf q},\omega) + F_2({\bf q}) G_1^{(n)}({\bf q},\omega) = \langle g_2^{(n)} \rangle. 
\label{GreenEq}
\end{multline}
Here $g_1^{(n)}=[S^+,(S^-)^n(S^+)^{n-1}]$ and $g_2^{(n)}=[S^-,(S^-)^n(S^+)^{n-1}]$ are functions depending on the spin operators, whose explicit form for $S=3/2$ ($n=1-3$) is given in Appendix \ref{appendix2}. Using the spectral theorem \cite{Zubarev1960}, we obtain the following equations:
\begin{equation}
\begin{split}
      \langle (S^-)^n (S^+)^n \rangle \quad \quad \quad  \quad \quad \quad \quad \quad \quad  \quad \quad \quad \quad \quad \quad \quad \quad \quad \\ 
    = \frac{1}{N}\sum_{\bf k}i\int \frac{d \omega}{2\pi}\frac{G_1^{(n)}({\bf q},\omega+i0^+) - G_1^{(n)}({\bf q},\omega-i0^-)}{e^{\beta \omega}-1} \\
        \langle (S^-)^n (S^+)^{n-1} S^- \rangle \quad \quad \quad  \quad \quad \quad \quad \quad \quad  \quad \quad \quad \quad \quad \quad  \quad \\ 
        = \frac{1}{N}\sum_{\bf k}i\int \frac{d \omega}{2\pi}\frac{G_2^{(n)}({\bf q},\omega+i0^+) - G_2^{(n)}({\bf q},\omega-i0^-)}{e^{\beta \omega}-1}.
\end{split}
\label{SpecT}
\end{equation}
The explicit form of the expressions in the left and right hand sides for $S=3/2$ is 
provided in Appendix \ref{appendix2}. One can notice that both sides of Eq.~(\ref{SpecT}) depend on the two kinds of variables: $\langle (S^y)^n \rangle$ and $\langle 
(S^-)^2(S^y)^{n-1} \rangle$. Therefore, we need to solve $4S=6$ equations self-consistently for each desired temperature to obtain the temperature-dependent magnetization $\langle S^y \rangle$.

\begin{figure}[tbp]
\centering
\includegraphics[width=1.0\linewidth]{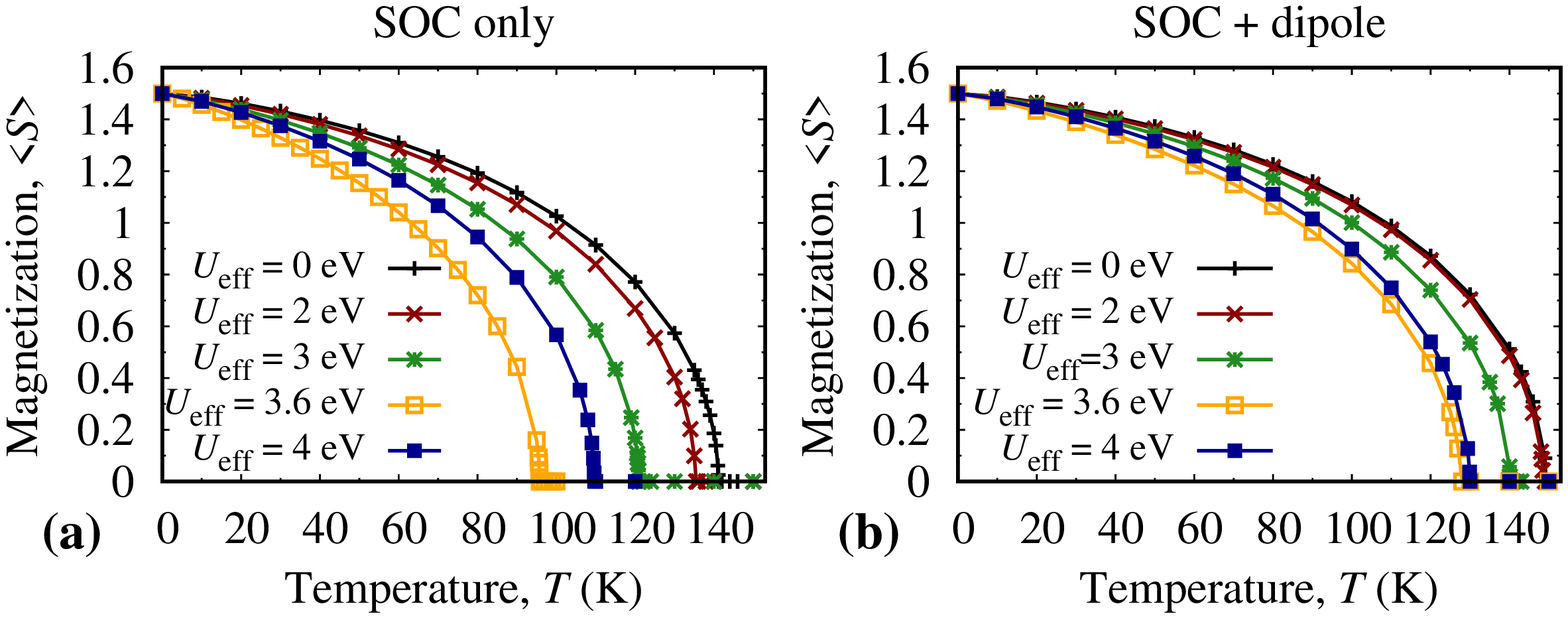}
\caption{Zero-field magnetization shown as a function of temperature calculated for different $U_\text{eff}$ parameters in ML-CrSBr with (left) and without (right) dipole interactions taken into account. The results are shown for the case of $U_\text{eff}$-dependent lattice constants.}
\label{magnetization}
\end{figure}

In Fig.~\ref{magnetization}, we show the calculated zero-fied magnetization as a function of temperature for different $U_\text{eff}$ values. The magnetization curve exhibits a typical shape, which allows us to determine the Curie temperature $T_C$ for each $U_\text{eff}$. At $T=0$, $\langle S^y \rangle = 1.5$, meaning that no quantum spin contraction effects, relevant for instance for highly anisotropic $S=1/2$ systems, are expected in ML-CrSBr. One can see that $T_C$ depends considerably on $U_\text{eff}$ in the situation when the dipolar interactions are neglected [Fig.~\ref{magnetization}(a)]. In this case, $T_C$ ranges from $\sim$140 K ($U_\text{eff}=0$) to $\sim$95 K ($U_\text{eff}=3.6$ eV). The lowest critical temperature corresponds to the minimum of the magnon gap [cf. Fig.~\ref{magnon_dispersion}(c)] attributed to the easy-plane instability. The presence of the dipolar interactions suppresses this instability, leading to a moderate variation of $T_C$ upon the change of $U_\text{eff}$. Specifically, we observe a small decrease of $T_C$ from 150 K to 130 K when $U_\text{eff}$ increases from 0 to 3.6 eV. As it was discussed earlier, this effect is mainly attributed to the behavior of the magnetic anisotropy, which becomes smaller for larger $U_\text{eff}$.

\begin{figure}[tb]
\centering
\includegraphics[width=1.0\linewidth]{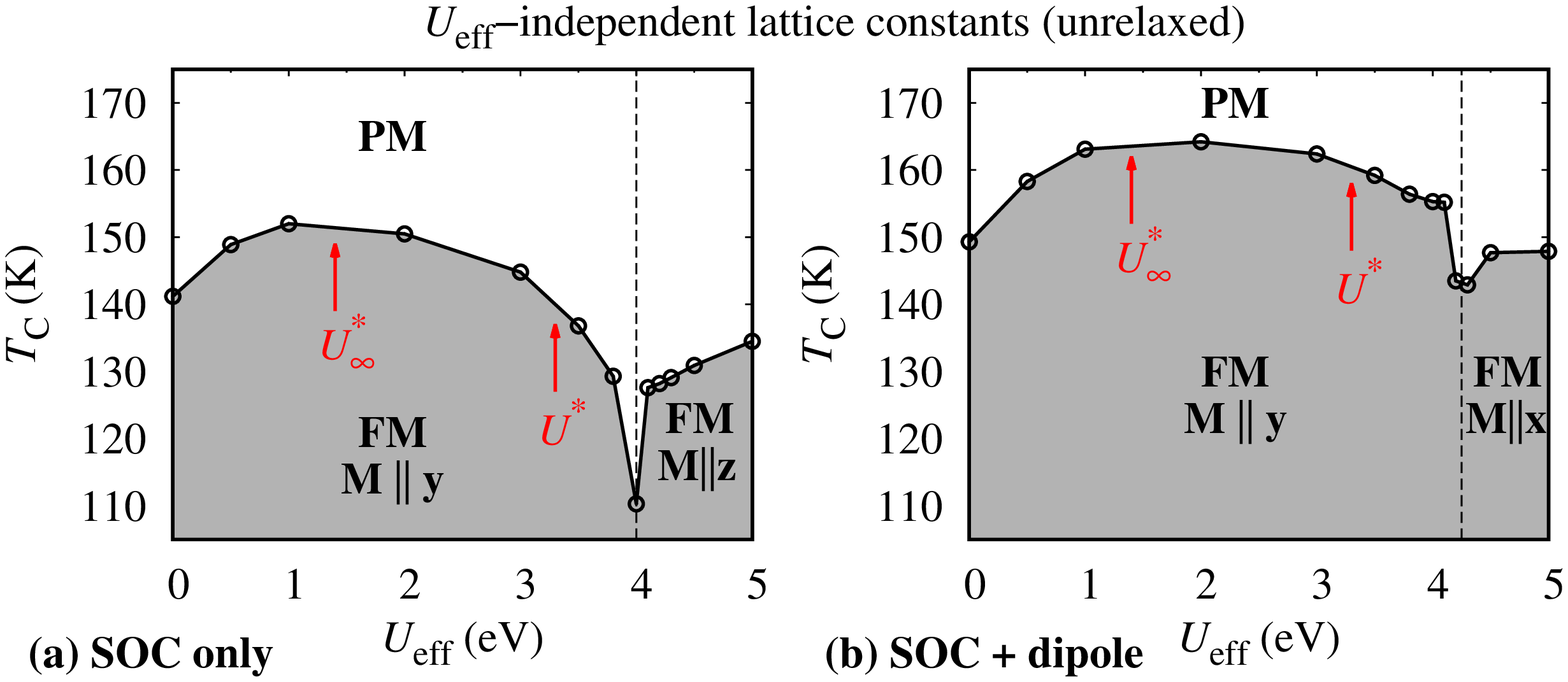}
\includegraphics[width=1.0\linewidth]{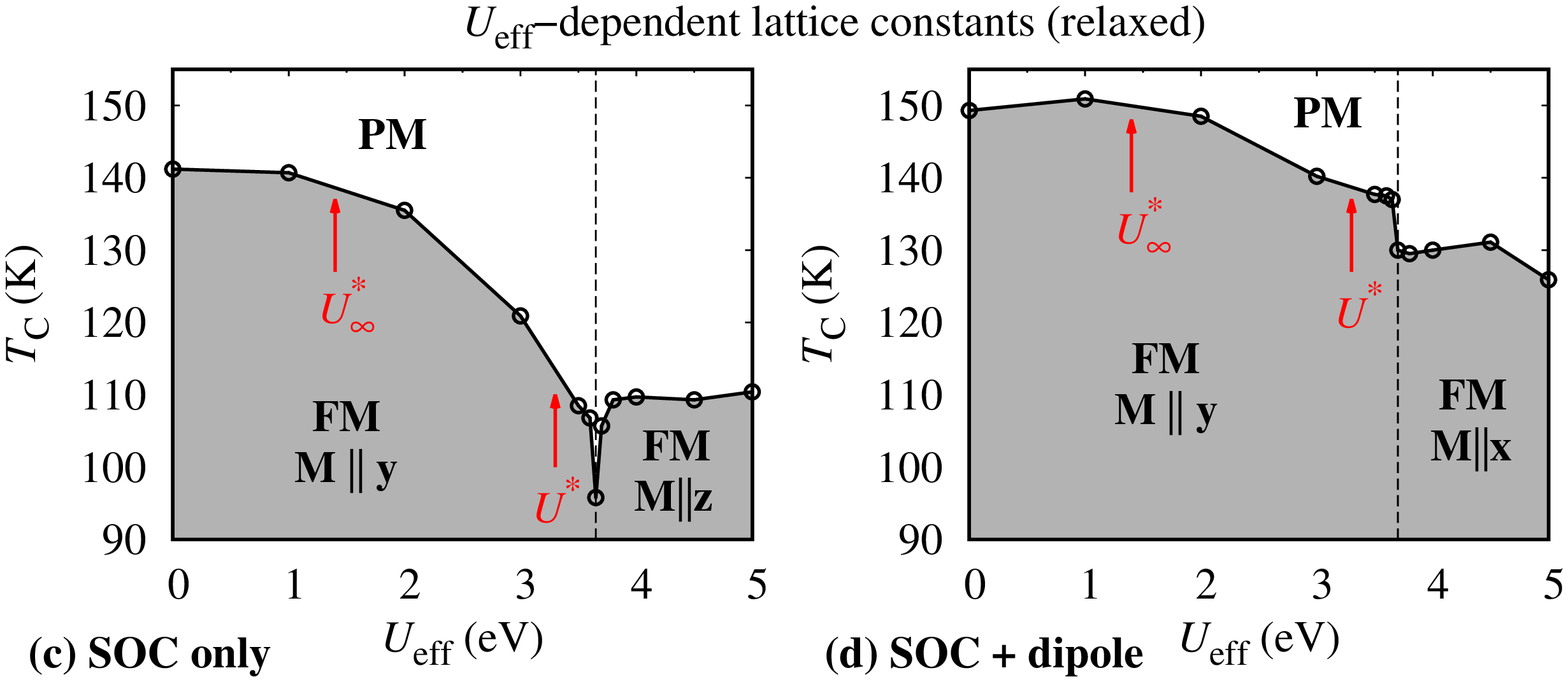}
\caption{The Curie temperature calculated in ML-CrSBr as a function of $U_\text{eff}$. Upper panels (a) and (b) show the results obtained without structural relaxation, while the results in the lower panels (c) and (d) do take $U_\text{eff}$-dependent relaxation into account. The results shown in the left panels (a) and (c) are obtained without the dipolar interaction, while in the right panels (b) and (d) this effect is included. The vertical dashed line corresponds to a point at which the easy axis changes its direction. The red arrows point to the effective Coulomb interaction estimated in Sec.~\ref{sec3b} for free-standing ($U^*$) and highly dielectically screened ($U^*_{\infty}$) ML-CrSBr.}
\label{Tc}
\end{figure}

In Fig.~\ref{Tc}, we present a summary of our findings showing the Curie temperature as a function of $U_\text{eff}$ for four different situations, i.e. depending on the presence of the dipolar interaction and of the relaxation effects. Although the ground magnetic state as well as the thermodynamical properties of ML-CrSBr are sensitive to the Coulomb interaction, its realistic values do not allow us to expect that the magnetic properties of ML-CrSBr can be efficiently manipulated by the dielectric screening. Indeed, the estimated strength of the Coulomb interaction for free-standing ML-CrSBr is around $U^* \approx 3.3$ eV, which is slightly smaller than the easy-plane instability point marked by the dash vertical lines in Fig.~\ref{Tc}. Therefore, we expect that the easy axis of ML-CrSBr is always pointing along the $y$-axis, irrespective of the external dielectric screening. Overall, the stability of the ferromagnetic phase could be somewhat increased in the presence of the dielectric environment. 
Provided that the dipolar interaction is always present under realistic conditions, the degree of the environment effects would depend on the structural details of real ML-CrSBr samples.
It is worth noting that the experimentally determined Curie temperature for ML-CrSBr (146 K \cite{Lee2021}) is in-between our estimates obtained for free-standing ML-CrSBr with relaxed ($\sim$140 K) and unrelaxed ($\sim$160 K) geometries.\\
\newline

\section{Conclusions\label{conclusion}}
We performed a systematic study of the isotropic and anisotropic magnetic properties in orthorhombic monolayer ferromagnetic CrSBr focusing on the effects of Coulomb interactions and their dielectric screening. We used the model of localized spins with a Hamiltonian including both the out-of-plane and in-plane magnetoctystalline anisotropy terms, as well as the magnetic dipole-dipole interactions.
The analysis of the thermodynamical properties is performed within the Green's functions formalism based on Tyablikov-like approximations for the spin operator decoupling. 

Despite highly anisotropic crystal structure and the electronic properties of ML-CrSBr, the exchange interactions are found to be weakly dependent on the crystallographic direction, resulting in an almost isotropic magnon propagation, which only slightly depends on the dielectric screening.
On the other hand, the magnetic anisotropy in CrSBr, predominantly originating from the single-ion anisotropy, is found to be very sensitive to the Coulomb interaction, vanishing at some point $U_{\mathrm{crit}}$ corresponding to the easy-plane instability. This point is unlikely to reach under realistic conditions because the effective Coulomb interaction of free-standing ML-CrSBr $U^* < U_{\mathrm{crit}}$. In the regime $U<U^*$, we find that the dipolar interaction plays no significant role, slightly increasing the magnon gap as well as the Curie temperature. The estimated Curie temperature of free-standing CrSBr is found to be around 140--160 K, depending on the structural details, which is in good agreement with the available experimental data. This value is expected to be slightly (not more than 10\%) larger for ML-CrSBr supported on dielectric substrates.

Our findings demonstrate a fundamentally different way for manipulating the magnetic properties of 2D magnets based on the environment-dependent strength of the magnetic anisotropy. This approach turns out to be limited for ML-CrSBr, where the magnetic properties do not vary within a wide range. Nevertheless, we expect that our result will stimulate further activities in this direction, expanding the spectrum of 2D magnets in which the proposed effects could be more efficient. 

\begin{acknowledgements}
The work was supported by European Research Council via Synergy Grant 854843 - FASTCORR.
\end{acknowledgements}

\bibliography{ref}

\appendix
\section{\label{appendix1}Energies of magnetic configurations}
\begin{figure}[tbp]
\centering
\includegraphics[width=0.95\linewidth]{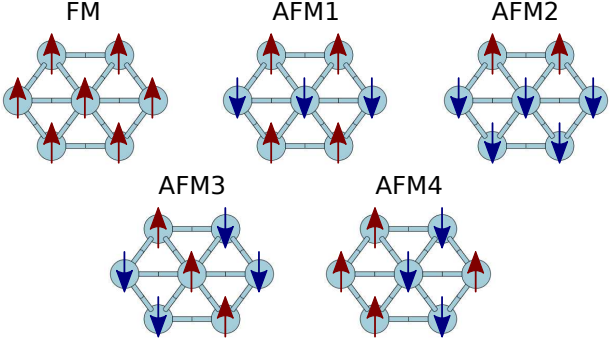}
\caption{Collinear magnetic configurations used to determine the parameters of the spin Hamiltonian.}
\label{configs}
\end{figure}

Here, we provide explicit expressions for the energies of the magnetic configurations used to estimate the parameters of the spin Hamiltonian. Figure~\ref{configs} shows five collinear FM and AFM configurations on a ($2 \times 2$) supercell for one particular magnetization direction. By changing the magnetization direction between $z$, $y$, and $x$, one obtains from Eqs.~(\ref{H0}), (\ref{HSIA}), and (\ref{HAE}) the following 15 equations for the magnetic energies per spin:

\begin{equation}    
\frac{E^x_{\mathrm{FM}}}{S^2} = 2(J_1+\Gamma_1) + 2(J_2+\Gamma_2) + 4(J_3 + \Gamma_3) + 4(J_4+\Gamma_4) + E 
\end{equation}

\begin{equation}
\frac{E^y_{\mathrm{FM}}}{S^2} = 2(J_1-\Gamma_1) + 2(J_2-\Gamma_2) + 4(J_3-\Gamma_3)  + 4(J_4-\Gamma_4) -E
\end{equation}

\begin{equation}
\frac{E^z_{\mathrm{FM}}}{S^2} = 2(J_1+K_1) + 2(J_2+K_2) + 4(J_3+K_3) + 4(J_4+K_4) + D
\end{equation}

\begin{equation}
\frac{E^x_{\mathrm{AFM1}}}{S^2} = 2(J_1+\Gamma_1) + 2(J_2+\Gamma_2) - 4(J_3+\Gamma_3) + 4(J_4+\Gamma_4) + E
\end{equation}

\begin{equation}
\frac{E^y_{\mathrm{AFM1}}}{S^2} = 2(J_1-\Gamma_1) + 2(J_2-\Gamma_2) - 4(J_3-\Gamma_3) + 4(J_4-\Gamma_4) -E
\end{equation}

\begin{equation}
\frac{E^z_{\mathrm{AFM1}}}{S^2} = 2(J_1+K_1) + 2(J_2+K_2) - 4(J_3+K_3) + 4(J_4+K_4) + D
\end{equation}

\begin{equation}
\frac{E^x_{\mathrm{AFM2}}}{S^2} = 2(J_1+\Gamma_1) - 2(J_2+\Gamma_2) - 4(J_4+\Gamma_4) + E\\
\end{equation}
\begin{equation}
\frac{E^y_{\mathrm{AFM2}}}{S^2} = 2(J_1-\Gamma_1) - 2(J_2-\Gamma_2) - 4(J_4-\Gamma_4) -E\\
\end{equation}

\begin{equation}
\frac{E^z_{\mathrm{AFM2}}}{S^2} = 2(J_1+K_1) - 2(J_2+K_2) - 4(J_4+K_4) + D\\
\end{equation}

\begin{equation}
\frac{E^x_{\mathrm{AFM3}}}{S^2} = -2(J_1+\Gamma_1) - 2(J_2+\Gamma_2) + 4(J_4+\Gamma_4) + E\\
\end{equation}
\begin{equation}
\frac{E^y_{\mathrm{AFM3}}}{S^2} = -2(J_1-\Gamma_1) - 2(J_2-\Gamma_2) + 4(J_4-\Gamma_4) -E\\
\end{equation}
\begin{equation}
\frac{E^z_{\mathrm{AFM3}}}{S^2} = -2(J_1+K_1) - 2(J_2+K_2) + 4(J_4+K_4) + D\\
\end{equation}

\begin{equation}
\frac{E^x_{\mathrm{AFM4}}}{S^2} = -2(J_1+\Gamma_1) - 2(J_2+\Gamma_2) - 4(J_4+\Gamma_4) + E\\
\end{equation}
\begin{equation}
\frac{E^y_{\mathrm{AFM4}}}{S^2} = -2(J_1-\Gamma_1) - 2(J_2-\Gamma_2) - 4(J_4-\Gamma_4) -E\\
\end{equation}
\begin{equation}
\frac{E^z_{\mathrm{AFM4}}}{S^2} = -2(J_1+K_1) - 2(J_2+K_2) - 4(J_4+K_4) + D.\\
\end{equation}
Here, the subscript $i$ in $J_i$, $\Gamma_i$, and $K_i$ corresponds to the notation given in Fig.~\ref{exchange}(b). The spin $S$ is assumed to be 3/2, in accordance with the Cr magnetic moment of 3.0 $\mu_B$ in CrSBr. By calculating the energy difference between difference configurations using DFT and solving the system of equations given above, we obtain 14 independent parameters, namely, $J_i$, $\Gamma_i$, $K_i$ ($i=1..4$), $D$, and $E$.

\section{\label{appendix2}Explicit form of 
the correlation functions given in the main text}
For $S=3/2$, Eq.~(\ref{SpecT}) determines six coupled equations containing six following variables: $\langle S^y \rangle$, $\langle (S^y)^2 \rangle$, $\langle (S^y)^3 \rangle$, $\langle (S^-)^2 \rangle$, $\langle (S^-)^2S^y \rangle$, and $\langle (S^-)^2(S^y)^2 \rangle$. The function $g_1^{(n)}=[S^+,(S^-)^n(S^+)^{n-1}]$ appearing in the right hand side of Eq.~(\ref{GreenEq}) can be expressed via these variables as
\begin{equation}
    g_1^{(n)} = (2nS^y + n^2 - n) \Phi^{(n-1)}(S^y), 
\end{equation}
where $\Phi^{(n)}(S^y)=(S^-)^n(S^+)^n$ or
\begin{equation}
\begin{split}
       \Phi^{(n)}(S^y) =  \prod_{p=1}^{n}\left[S(S+1) - (n-p)(n-p+1) \right. \\ \left. - (2n-2p+1)S^y - (S^y)^2\right].
\end{split}
\end{equation}
Similarly, for $g_2^{(n)}(S^-,S^z) = [S^-,(S^-)^n(S^+)^{n-1}]$ we have
\begin{equation}
\begin{split}
  g_2^{(n)} =  -\left[ (n-1)(n-2)(S^-)^2 \right. \\ \left. + 2(n-1)(S^-)^2S^y \right]\Phi^{(n-2)}(S^y).
\end{split}
\end{equation}
For $n=1-3$, i.e. $S=3/2$, we obtain the following explicit expressions:
\begin{align*}
\notag
\langle g_1^{(1)} \rangle &= 2\x \\
\langle g_1^{(2)} \rangle &= \frac{15}{2} + 13\x - 6\y - 4\z \\
\langle g_1^{(3)} \rangle &= \frac{45}{2} - 42 \x - 18\y + 24\z \\
\langle g_2^{(1)} \rangle &= 0 \\
\langle g_2^{(2)} \rangle &= -2 \langle (S^{-})^2S^y \rangle \\
\langle g_2^{(3)} \rangle &= -9\langle(S^{-})^2\rangle - 12\langle(S^{-})^2S^y\rangle + 12\langle(S^{-})^2(S^y)^2\rangle.
        \notag
\end{align*}

The correlation functions appearing in the left hand side of Eq.~(\ref{SpecT}) can be expressed in a similar manner as follows:
\begin{align*}
	&\langle (S^-)(S^+) \rangle = \frac{15}{4} - \x - \y \\    
	&\langle (S^-)^2(S^+)^2 \rangle = 6 - 13\x + 4\z \\    
	&\langle (S^-)^3(S^+)^3 \rangle = -\frac{9}{4} + \frac{3}{2}\x + 9\y - 6\z \\
	&\langle (S^-)^1(S^+)^0S^- \rangle = \langle (S^-)^2 \rangle  \\
	&\langle (S^-)^2(S^+)^{1}S^- \rangle = \frac{15}{4} \langle (S^-)^2 \rangle  + \langle (S^-)^2S^y \rangle  - \langle (S^-)^2(S^y)^2 \rangle \\
	&\langle (S^-)^3(S^+)^{2}S^- \rangle = \frac{27}{2} \langle (S^-)^2 \rangle  - 6\langle (S^-)^2(S^y)^2 \rangle.
    \notag
\end{align*}

A more convenient form of Eq.~(\ref{SpecT}) reads
\begin{equation}
    \begin{split}
        \langle (S^-)^n (S^+)^n \rangle  = \frac{1}{N}\sum_{\bf q}\frac{1}{2E^{\pm}({\bf q})[e^{\beta E({\bf q})}-1]} \\ \times \{ F_2({\bf q})\langle g_2^{(n)}\rangle + \langle g_1^{(n)} \rangle[E^{\pm}({\bf q}) + F_1({\bf q})] \rangle \} \\
        - \frac{1}{N}\sum_{\bf q}\frac{1}{2E^{\pm}({\bf q})[e^{-\beta E({\bf q})}-1]} \\
         \times \{ F_2({\bf q})\langle g_2^{(n)}\rangle - \langle g_1^{(n)} \rangle[E^{\pm}({\bf q}) - F_1({\bf q})] \rangle \},
    \end{split}
\end{equation}

\begin{equation}
    \begin{split}
        \langle (S^-)^n (S^+)^n S^- \rangle  = -\frac{1}{N}\sum_{\bf q}\frac{1}{2E^{\pm}({\bf q})[e^{\beta E({\bf q})}-1]} \\ \times \{ F_2({\bf q})\langle g_1^{(n)}\rangle - \langle g_2^{(n)} \rangle[E^{\pm}({\bf q}) - F_1({\bf q})] \rangle \} \\
        + \frac{1}{N}\sum_{\bf q}\frac{1}{2E^{\pm}({\bf q})[e^{-\beta E({\bf q})}-1]} \\
         \times \{ F_2({\bf q})\langle g_1^{(n)}\rangle + \langle g_2^{(n)} \rangle[E^{\pm}({\bf q}) + F_1({\bf q})] \rangle \}.
    \end{split}
\end{equation}
In order to solve these equations, we use an in-house developed code.

\end{document}